\documentclass[onecolumn,nofootinbib,superscriptaddress,notitlepage,amsmath,amssymb,fleqn,aps,prd,preprintnumbers,10pt]{revtex4-1}
\usepackage{xcolor,braket,graphicx,array,xspace,wasysym,placeins,hyperref,todonotes,comment,cancel}
\usepackage[all]{hypcap}

\newcommand{\eps}{\varepsilon}

\newcommand{\dilog}{\operatorname{Li_2}}
\newcommand{\re}{\operatorname{Re}}
\newcommand{\atan}{\operatorname{atan}}

\newcommand{\MCatNLO}{M\protect\scalebox{0.8}{C}@N\protect\scalebox{0.8}{LO}\xspace}

\newcommand{\Alaric}{A\scalebox{0.8}{LARIC}\xspace}
\newcommand{\Sherpa}{S\scalebox{0.8}{HERPA}\xspace}

\hypersetup{hidelinks,
  pdfauthor={Stefan Hoeche,Daniel Reichelt},
  pdftitle={Resonance- and Width-aware Parton Shower Evolution and NLO Matching},
  pdfkeywords={QCD, Perturbation Theory}
}

\begin{document}
\preprint{FERMILAB-PUB-26-0230-T, CERN-TH-2026-085, MCNET-26-06}
\title{Resonance- and Width-aware Parton Shower Evolution and NLO Matching}
\author{Stefan H{\"o}che}
\affiliation{Fermi National Accelerator Laboratory, Batavia, IL, 60510, USA}
\author{Daniel Reichelt}
\affiliation{Theoretical Physics Department, CERN, CH-1211 Geneva, Switzerland}

\begin{abstract}
We introduce a technique for the next-to-leading order accurate simulation of
$e^+e^-\to W^+W^-b\bar{b}$ that respects the resonant nature of the process above and
near the top-quark pair production threshold. The parton-shower evolution, infrared
subtraction and NLO matching account in particular for finite width effects beyond the
Breit-Wigner structure considered in resonance-aware approaches. We present first
phenomenological results relevant to a potential future electron-positron collider
and provide a publicly available simulator based on the \Alaric parton shower
and the \Sherpa event generator.
\end{abstract}

\maketitle

\section{Introduction}
The top quark plays a special role in high-energy particle phenomenology. 
With a Yukawa coupling of order one, it affects the stability of the electroweak
vacuum~\cite{Cabibbo:1979ay,Hung:1979dn,Lindner:1985uk,Sher:1988mj,Alekhin:2012py,Degrassi:2012ry},
making precision measurements of its properties one of the most important goals of current
and future collider experiments. It is expected that electroweak parameters extracted at
a lepton-lepton Future Circular Collider (FCC-ee)~\cite{FCC:2025lpp,FCC:2025uan,FCC:2025jtd}
will require knowledge of the top-quark mass to the level of 50~MeV~\cite{Narain:2022qud},
a precision requirement that can likely only be met at the FCC-ee itself. Intriguingly,
top-quark final states have never before been observed at a lepton collider.
Even at low statistics, the precision goals mandate a significant improvement in the
modeling of this process compared to hadron colliders. This is particularly important
in the context of particle-level Monte-Carlo event simulations~\cite{Campbell:2022qmc}. 

With the top quark being the only quark to decay before hadronization, its QCD
interactions can always be computed using perturbative methods. In the context
of event generators, the radiative corrections are typically approximated
and resummed by parton showers. However, due to the finite width of the top quark,
a traditional parton-shower approach is insufficient. On the one hand, the dipole
structure of gluon radiation originating from the decay products of the top quarks must be
modeled correctly, taking into account color-correlations with other partons in the final state.
On the other hand, the fact that the top-quark production process is resonant must be accounted for.
Satisfying both of these requirements at the same time is quite challenging. In fact they seem
to be mutually exclusive at first, because most modern parton showers achieve a correct simulation
of the eikonal radiation pattern with the help of phase-space sectorization or partial
fractioning of the matrix element. In these approaches, the recoil that emerges when the
splitting parton is shifted off mass shell is distributed
onto the color partner or the complete color multipole. In the case of a
$W^+W^-b\bar{b}$ final state, this would imply that the recoil from simulated
gluon radiation off the $b$ ($\bar{b}$) quark would be compensated (at least partially)
by the $\bar{b}$ ($b$) quark. The immediate consequence is that QCD radiative corrections
would alter the virtuality of the intermediate resonances. This can change the rate of
the underlying Born process quite dramatically, and create unintended consequences if
one assumes that the parton-shower evolution is  truly Markovian in nature. A better
approach is therefore to model the radiation with a technique that does not affect
the virtuality of the intermediate top quarks.

A general algorithm, based on a resonance-aware partitioning of the radiative phase space
and suitable for next-to-leading order (NLO) matching of parton showers was proposed
in Refs.~\cite{Jezo:2015aia,Jezo:2016ujg}. It has been successfully applied to multiple
problems in the context of Large Hadron Collider physics~\cite{FerrarioRavasio:2018whr,
  Jezo:2023rht,Denner:2026ztd}. However, it is still insufficient for high-precision
top-quark measurements, because finite width effects are taken into account only in the
identification of the intermediate resonances. We therefore propose a different approach
to the problem, which is needed in particular to satisfy the much higher precision
requirements of an FCC-ee. Our technique is partially based on a method previously
applied to fixed order NLO calculations~\cite{Hoche:2018ouj} and makes use of the
identified particle subtraction algorithm of Catani and Seymour~\cite{Catani:1996vz}
to retain the virtuality of intermediate resonances. We extend it to include the known
finite width effects near threshold, which were first analyzed in~\cite{Jikia:1990qm,
  Khoze:1992rq,Dokshitzer:1992nh,Orr:1994ar}. We use the \Alaric parton shower~\cite{
  Herren:2022jej,Assi:2023rbu,Hoche:2024dee,Hoche:2025gsb,Hoche:2025anb} for resummation
and the \Sherpa event generation framework~\cite{Gleisberg:2003xi,Gleisberg:2008ta,
  Sherpa:2019gpd,Sherpa:2024mfk} for the corresponding next-to-leading order calculations.
The infrared subtraction terms are computed analytically.
As a first phenomenological application, we employ the new technique to
predict top-quark pair production at a hypothetical FCC-ee at $\sqrt{s}=365$~GeV.

The manuscript is structured as follows: Sections~\ref{sec:resonance-aware}
and~\ref{sec:width-aware} discuss the problem of resonance- and width-aware
parton evolution in the \Alaric framework and present our solution based on the
identified particle subtraction method of Catani and Seymour.
Section~\ref{sec:nlo} introduces the corresponding technique for fixed-order
calculations and presents the new analytic expressions for the integrated
infrared subtraction terms. In Sec.~\ref{sec:pheno} we validate the
fixed-order NLO calculation and discuss first phenomenological results.
Section~\ref{sec:outlook} contains an outlook.

\section{Resonance-aware parton evolution in Alaric}
\label{sec:resonance-aware}
The infrared structure of the real-emission corrections to a process with $m$
partons at Born level can be described using the Catani-Seymour dipole factorization
formulae~\cite{Catani:1996vz,Catani:2002hc}~\footnote{Note that we consider all
  partons to be in the final state.}:
\begin{equation}\label{eq:cs_dipole_factorization}
    \,_{m+1}\langle 1,\ldots,i,\ldots,j,\ldots,m+1|1,\ldots,i,\ldots,j,\ldots,m+1\rangle_{m+1}
    =\sum_{k\neq a}\mathcal{D}_{ij,k}(p_1,\ldots,p_{m+1})\;,
\end{equation}
where the individual dipole contributions are given by
\begin{equation}
    \mathcal{D}_{ij,k}(p_1,\ldots,p_{m+1})
    =-\frac{1}{2p_ip_j}\,_m\langle\tilde{1},\ldots,\widetilde{ij},\ldots,\widetilde{m\!+\!1}|
    \frac{{\bf T}_i{\bf T}_k}{{\bf T}_i^2}V_{ij,k}
    |\tilde{1},\ldots,\widetilde{ij},\ldots,\widetilde{m\!+\!1}\rangle_m\;.
\end{equation}
The differential insertion operator $V_{ij,k}$ consists of a scalar term
and a pure splitting component~\cite{Campbell:2025lrs,Hoche:2025vto},
which will be labeled with superscripts $\,^{(s)}$ and $\,^{(p)}$, respectively
(see also Ref.~\cite{Hoche:2025gsb}).
\begin{equation}\label{eq:soft_coll_sinsertion_split}
    V_{ij,k}=V_{ij,k}^{(s)}+V_{ij,k}^{(p)}\;.
\end{equation}
In analogy to the differential insertion operator, $V_{ij,k}$, we define
the integral of its spin average over the one-emission phase space, $\mathbf{I}$,
as a sum of scalar and splitting components.
Following the notation of Ref.~\cite{Hoche:2025gsb}, we have
\begin{equation}\label{eq:soft_coll_iinsertion_split}
    \mathbf{I}=\int[{\rm d}p_j]\frac{1}{2p_ip_j}\langle V_{ij,k}\rangle=
    -\frac{\alpha_s}{2\pi} \frac{1}{\Gamma(1-\eps)}\left(\frac{4\pi\mu^2}{2p_ip_k}\right)^\eps
    \frac{\mathbf{T}_{ij} \mathbf{T}_{k}}{\mathbf{T}_{ij}^2}
    \left(I^{(s)}_{i,k}+I^{(p)}_{ij}\right)\;.
\end{equation}
The scalar components of
Eqs.~\eqref{eq:soft_coll_sinsertion_split} and~\eqref{eq:soft_coll_iinsertion_split}
only exist for splittings where the emitted parton with label $j$ is a gluon.
They are spin-independent, as they represent the semi-classical contribution
to the real radiative corrections~\cite{Gell-Mann:1954wra,Brown:1968dzy}.
In the \Alaric parton-shower algorithm, they are defined as the differential
and integrated form of the semi-classical eikonal, partial fractioned
in terms of relative angles in the frame defined by an auxiliary vector, $n$,
which is tied to the recoil definition~\cite{Catani:1996vz,Herren:2022jej},
cf.\ App.~\ref{app:kinematics}.
\begin{equation}\label{eq:eik_partfrac_alaric}
    V_{ig,k}^{(s)}(p_i,p_j,n)=8\pi\mu^{2\eps}\alpha_s\, C_i\,
    \frac{2(p_ip_k)(p_in)}{(p_ip_j)(p_kn)+(p_kp_j)(p_in)}\;.
\end{equation}

For a process with only one radiating QCD multipole, one would usually choose
the recoil vector to be the sum of all QCD charged particles in the multipole.
In an $e^+e^-\to W^+W^-b\bar{b}$ scattering, this choice corresponds to the
$b\bar{b}$ pair. Upon gluon radiation off one of the quarks, the recoil would
then be taken by the antiquark and vice versa. Let us assume that the $b$-quark
is the emitting parton. Using the notation of Ref.~\cite{Catani:1996vz}, labeling
momenta before radiation by $\tilde{p}$ and momenta after radiation by $p$,
the virtualities of the top (anti-)quark propagators will then change from
$(p_{W^+}+\tilde{p}_b)^2$ and $(p_{W^-}+\tilde{p}_{\bar{b}})^2$
to $(p_{W^+}+p_b+p_g)^2$ and $(p_{W^-}+p_{\bar{b}})^2$. The differential rate
for the underlying $e^+e^-\to W^+W^-b\bar{b}$ process changes approximately
by the ratio between the $t$ and $\bar{t}$ propagators before and after
the splitting:
\begin{equation}
  \frac{\big[(p_{W^+}+\tilde{p}_b)^2-m_t^2\big]^2+m_t^2\Gamma_t^2}{
    \big[(p_{W^+}+p_b+p_g)^2-m_t^2\big]^2+m_t^2\Gamma_t^2}
  \frac{\big[(p_{W^-}+\tilde{p}_{\bar{b}})^2-m_t^2\big]^2+m_t^2\Gamma_t^2}{
    \big[(p_{W^-}+p_{\bar{b}})^2-m_t^2\big]^2+m_t^2\Gamma_t^2}
\end{equation}
When the transverse momentum of the gluon relative to the parent $b$-quark
is of order of the top-quark width, this factor can induce changes of $\mathcal{O}(1)$,
independent of the hardness of the emitting $b$-quark. This contradicts
the assumption that the recoil effects on the hard process are sub-leading
power corrections to the resummed result. Parton-shower resummation using
a standard recoil scheme is therefore not a sensible approach in this case.
The problem was recognized in the context of NLO parton-shower matching
in Ref.~\cite{Jezo:2015aia}, and a solution was proposed that relies on
a recoil scheme which leaves the virtuality of the resonance unaffected.
This technique is referred to as resonance aware evolution and matching.

The \Alaric parton shower provides a natural way to satisfy the requirements
for resonance aware evolution. We first note that the choice of the auxiliary
vector $n$ in Eq.~\eqref{eq:eik_partfrac_alaric} is in principle arbitrary,
as long as it is guaranteed to provide a hard scale~\cite{Herren:2022jej}.
We can use this ambiguity, and the fact that $n$ is defined in terms of the
recoil vector, $K=n-p_j$, to construct a matching scheme that preserves
the virtuality of the identified resonances. In the case of the
$e^+e^-\to W^+W^-b\bar{b}$ process, it is applied as follows: If a parton
originating in the decay of the (anti-)top quark splits, we set the recoil
vector to the sum of the other (anti-)top decay products~\cite{Hoche:2018ouj}.
This ensures that the necessary momentum reshuffling happens entirely
within the decay products of the identified resonance. However, 
an additional complication arises because the resonances are color connected.
The treatment of eikonal factors in Ref.~\cite{Herren:2022jej,Hoche:2025gsb}
assumes that for radiation off both partons forming a dipole, the same recoil
vector, $\tilde{K}$, will be chosen. Only then can we use the scalar emission kernel
in Eq.~\eqref{eq:eik_partfrac_alaric}, which depends on the recoil vector
$\tilde{K}$ through $n$. The technique fails for resonance aware evolution of
dipoles that are split across different resonances, because they are now assigned
different recoil vectors. The solution is to replace Eq.~\eqref{eq:eik_partfrac_alaric}
by the original scalar emission operator of Catani and Seymour, which is based on
a partial fractioning in terms of scalar invariants, and therefore independent
of $\tilde{K}$:
\begin{equation}\label{eq:eik_partfrac_cs}
    V_{ig,k}^{(s,{\rm CS})}(p_i,p_j)=8\pi\mu^{2\eps}\alpha_s\, C_i\,
    \frac{2(p_ip_k)}{(p_ip_j)+(p_kp_j)}~.
\end{equation}
During the subsequent parton shower evolution, each newly created particle
inherits the assignment to a particular resonance from their emitter.
The recoil from additional emissions is thus always distributed among
the particles identified as originating in the same resonance.
An additional complication arises, however, in the matching to next-to-leading order
calculations. In order to define the initial conditions of the parton shower
for events with a real emission (called H-events in the \MCatNLO literature)
one must assign a unique parton-shower branching history to the higher-multiplicity
final state. This can be achieved consistently with the help of well-known
algorithms for multi-jet merging~\cite{Andre:1997vh}. More specifically, one can
perform a single clustering step, with the branching selected probabilistically, according
to the parton shower evolution kernels. In the case of $e^+e^-\to W^+W^-b\bar{b}$,
this assigns the additional gluon to either the $b$ or $\bar{b}$ quark, which
in turn determines its recoil vector. 

The assignment of recoil, and the change from
Eq.~\eqref{eq:eik_partfrac_alaric} to Eq.~\eqref{eq:eik_partfrac_cs},
define the resonance aware evolution scheme in the \Alaric formalism.
No changes are needed for the splitting operator, $V^{(p)}_{ij,k}$,
in Eq.~\eqref{eq:soft_coll_sinsertion_split}.

\section{Width-aware parton evolution close to the \texorpdfstring{\boldmath$t\bar{t}$}{tt} threshold}
\label{sec:width-aware}
In this section we will introduce the width aware parton evolution, which
allows us to significantly increase the precision of the simulation of processes
with identified resonances close to the threshold. Consider again the process
$e^+e^-\to W^+W^-b\bar{b}$. We label the momenta of the intermediate, potentially
off-shell top quark and anti-top quark as $q_i$ and $q_k$, and the momenta of the
final-state (massless) b-quark and anti-b quark as $p_i$ and $p_k$. Assuming a
single additional gluon in the final state, we label its momentum as $p_j$.
Following Refs.~\cite{Jikia:1990qm,Khoze:1992rq}, the matrix element squared 
above and close to threshold receives contributions from the dipoles formed by
the top and bottom quarks (i.e.\ from the individual top decay processes)
\begin{equation}
  \begin{aligned}
    B_i^2 &= \frac{2 p_i q_i}{(p_i p_j)(p_j q_i)} - \frac{m_t^2}{(p_j q_i)^2} &&=
    \frac{1}{p_ip_j}\frac{2 p_i q_i}{(p_i p_j)+(p_j q_i)} +
    \frac{1}{q_ip_j}\frac{2 p_i q_i}{(p_i p_j)+(p_j q_i)} - \frac{m_t^2}{(q_i p_j)^2}\;,\\
    B_k^2 &= \frac{2 p_k q_k}{(p_k p_j)(p_j q_k)} - \frac{m_t^2}{(p_j q_k)^2} &&=
    \frac{1}{p_kp_j}\frac{2 p_k q_i}{(p_k p_j)+(p_j q_k)} +
    \frac{1}{q_kp_j}\frac{2 p_k q_i}{(p_k p_j)+(p_j q_k)} - \frac{m_t^2}{(q_k p_j)^2}\;.\\
  \end{aligned}
\end{equation}
There is also a width suppressed interference contribution,
\begin{equation}
  \begin{aligned}
    2 \re[B_i B_k^*] =&\; 
      \frac{2 m_t^2 \Gamma_t^2 ((p_j q_i)(p_j q_k) + m_t^2
      \Gamma_t^2)}{((p_j q_i)^2 + m_t^2 \Gamma_t^2)((p_j q_k)^2 + m_t^2 \Gamma_t^2)}\\
      &\;\times\left(\frac{q_i q_k}{(q_i p_j)(p_j q_k)}
      - \frac{p_i q_k}{(p_ip_j)(p_j q_k)}
      + \frac{p_i p_k}{(p_i p_j)(p_k p_j)}
      - \frac{p_k q_i}{(p_k p_j)(p_j q_i)}\right)
  \end{aligned}
\end{equation}
as well as additional contributions from the top-antitop dipole that are suppressed
by the top quark velocity $v_t$. For center of mass energies close to the threshold
(but not so close that bound-state effects are important), we can neglect the
velocity-suppressed terms and substitute
\begin{equation}
    p_j q_i \to E_j m_t\;,\qquad
    q_1 q_2 \to m_t^2\;,\qquad
    \frac{m_t^2 \Gamma_t^2 ((p_j q_i)(p_j q_k) + m_t^2
      \Gamma_t^2)}{((p_j q_i)^2 + m_t^2 \Gamma_t^2)((p_j q_k)^2 + m_t^2
      \Gamma_t^2)} \to \frac{\Gamma_t^2}{E_j^2 + \Gamma_t^2}\;.
\end{equation}
Collecting all terms that are enhanced for emissions collinear to $p_i$, we find
\begin{equation}
  \begin{aligned}
  &\sim \frac{1}{p_ip_j}\frac{2 p_i q_i}{(p_i p_j)+(p_j q_i)} +
  \frac{\Gamma_t^2}{E_j^2 + \Gamma_t^2} \left(\frac{1}{p_ip_j}\frac{2p_i p_k}{(p_i p_j)+(p_k
        p_j)} - \frac{1}{p_ip_j}\frac{2p_i q_k}{(p_i
    p_j)+(p_j q_k)}\right)\\
  &= \frac{1}{p_ip_j}\left[\frac{2p_i p_k}{(p_i p_j)+(p_k
    p_j)} + \frac{E_j^2}{E_j^2+\Gamma_t^2}\left(\frac{2 p_i q_i}{(p_i p_j)+(p_j q_i)}-\frac{2p_i p_k}{(p_i p_j)+(p_k
    p_j)}\right) \right] + \mathcal{O}(v_t)\;.
  \end{aligned}
\end{equation}
Here we have also exploited an ambiguity in the threshold case. When both top quarks
are nearly at rest, we find $p_iq_k = p_iq_i+\mathcal{O}(v_t)$.
The energy of the gluon, $E_j$, is evaluated in the rest frame of the top quarks,
\begin{equation}
  q_i = p_W + p_i + p_j = \tilde{p}_W + \tilde{p}_i = n + p_i\;,
\end{equation}
and we approximate $p_jq_i = np_j + p_ip_j = (1-z)\tilde{p_i}\tilde{K} + p_ip_j \sim (1-z)\tilde{p}_i\tilde{K}$,
which gives
\begin{equation} \label{eq:chi}
        \frac{E_j^2}{E_j^2+\Gamma_t^2} = \frac{(q_ip_j)^2}{(q_ip_j)^2 + q_i^2\Gamma_t^2}
        \sim \frac{(1-z)^2}{(1-z)^2+\gamma^2}=:\chi(z)\;,
        \qquad\text{where}\qquad
        \gamma = \sqrt{q_i^2}\frac{\Gamma_t}{\tilde{p}_i\tilde{K}}
\end{equation}
Motivated by the above, we use the following scalar splitting kernel:
\begin{equation}\label{eq:soft-kernel}
  \begin{aligned}
    V_{ij,k}^{(s,\text{w})}(p_i,p_j,p_k,q_i)&=8\pi\mu^{2\eps}\alpha_s\, C_i\,
    \left[\frac{2(p_ip_k)}{(p_ip_j)+(p_kp_j)} +
      \chi(z)\left(\frac{2(p_iq_i)}{(p_ip_j)+(q_ip_j)}-\frac{2(p_ip_k)}{(p_ip_j)+(p_kp_j)}\right)
      \right]\\ 
    &= V_{ij,k}^{(s,\text{CS})}(p_i,p_j,p_k) + V_{ij,q_i}^{(s,\text{w})}(p_i,p_j,q_i)\;.
  \end{aligned}
\end{equation}
A parton shower with similarly modified evolution kernels was first studied in Ref.~\cite{Khoze:1994fu}\footnote{
  Note that the definition of $\chi(z)$ in Eq.~\eqref{eq:chi} corresponds to $1-\chi(\omega)$ in Ref.~\cite{Khoze:1994fu}.}.
Our main goal is to consistently include the changes in an NLO matched parton shower with NLL safe momentum mapping.

\section{Fixed-order infrared subtraction}
\label{sec:nlo}
In this section we derive the integrated infrared subtraction counterterms in
Eq.~\eqref{eq:soft_coll_iinsertion_split} that correspond to the scalar radiator
functions in Eqs.~\eqref{eq:eik_partfrac_cs} and~\eqref{eq:soft-kernel}.
We use the techniques from Refs.~\cite{Catani:1996vz,Assi:2023rbu} to perform the relevant
integrals. We define the Catani-Seymour result and the width-dependent,
finite remainder as
\begin{equation}\label{eq:soft_iinsertion_split}
    \mathbf{I}^{(s)}=\int[{\rm d}p_j]\frac{1}{2p_ip_j}
    \left(\langle V_{ij,k}^{s,\text{CS}}\rangle+ \langle V_{ij,q_i}^{s,\text{w}}\rangle\right)=
    -\frac{\alpha_s}{2\pi} \frac{1}{\Gamma(1-\eps)}\left(\frac{4\pi\mu^2}{2p_ip_k}\right)^\eps
    \frac{\mathbf{T}_{ij} \mathbf{T}_{k}}{\mathbf{T}_{ij}^2}
    \left(I^{(s,\text{CS})}_{i,k}+I^{(s,\text{w})}_{i,j}\right)\;.
\end{equation}
The first term in Eq.~\eqref{eq:soft-kernel} can be obtained from the results
in App.~\ref{app:soft-integrals} by setting $A=k$.
\begin{equation}
  \begin{aligned}
  I^{(s,\text{CS})}_{ik} &= \int_0^1 \mathop{\mathrm{d}z}
    \left(-\frac{\delta(1-z)}{2\eps}+\frac{z}{\left[1-z\right]_{+}}
    - 2\eps z \left[\frac{\log (1-z)}{1-z}\right]_{+} \right)\\
    &\phantom{= \int_0^1 \mathop{\mathrm{d}z} } \times \frac{2}{\pi} z^{-\eps}
    \left(\frac{np_k}{np_i} \frac{l_{ik}^2}{4}\right)^{\eps}\frac{\Gamma^2(1-\eps)}{\Gamma(1-2\eps)}\;
     l_il_{i+k}I^{(1)}_{1,1}(l_il_{i+k},l_{i+k}^2)~.
  \end{aligned}
\end{equation}
In terms of the variables defined in Ref.~\cite{Hoche:2025gsb},
\begin{equation}
    \rho \equiv \frac{2 \tilde{p}_k \tilde{K}}{2p_ip_k},\qquad
    \tau \equiv \frac{2 \tilde{p}_i \tilde{K}}{2p_ip_k},\qquad
    \mu_K \equiv \frac{\tilde{K}^2}{2p_ip_k}\;,
\end{equation}
we find
\begin{equation}
  I^{(s,\text{CS})}_{ik} = \frac{1}{\eps^2} + \frac{2}{\eps} +  6-\frac{\pi^2}{2} +
  4\left[\dilog\left(\frac{\rho-\tau+\sqrt{(\rho+\tau)^2-4\mu_K}}{2\tau}\right)
  + \dilog\left(\frac{\rho+\tau-\sqrt{(\rho+\tau)^2-4\mu_K}}{2\tau}\right)\right]\;,
\end{equation}
which agrees with Ref.~\cite{Catani:1996vz}. 
For the finite term, we obtain the difference of two integrands of the type
of App.~\ref{app:soft-integrals}, modulated by $\chi(z)$, which regulates
the soft divergence. The result is finite in $\eps$ and reads
\begin{equation}\label{eq:finite_remainder_unintegrated}
  \begin{aligned}
  I^{(s,\text{w})}_{ik} =&\; \int_0^1 \mathop{\mathrm{d}z}
    \frac{z^{1-\eps}}{(1-z)^{1+2\eps}} \frac{2}{\pi}
    \left(\frac{np_k}{np_i} \frac{l_{ik}^2}{4}\right)^{\eps}
    \frac{\Gamma^2(1-\eps)}{\Gamma(1-2\eps)}\frac{(1-z)^2}{(1-z)^2+\gamma^2}\\ 
     &\phantom{\int_0^1 \mathop{\mathrm{d}z} } \times \left[\,
     l_il_{i+q_i} I_{1,1}^{(1)}\left(l_il_{i+q_i},l_{i+q_i}^2\right)
     - l_il_{i+k}I^{(1)}_{1,1}\left(l_il_{i+k},l_{i+k}^2\right)\right] \\
     =&\;\int_0^1 \mathop{\mathrm{d}z} \frac{2z}{1-z} \frac{(1-z)^2}{(1-z)^2+\gamma^2}
     \left[\,\ln \frac{p_ip_k}{p_iq_i}
     + \ln \left(\frac{q_i^2}{z2p_iq_i}+1\right) \right] + \mathcal{O}(\eps)\;.
  \end{aligned}
\end{equation}
The $z$-integral for the $\eps^0$ coefficient can be performed analytically.
The result is
\begin{equation}\label{eq:finite_remainder}
    \begin{aligned}
        I^{(s,\text{w})}_{ik} =& -2 \ln \frac{(p_ip_k) q_i^2}{2(p_iq_i)^2}
        \left[\,1-\gamma\atan\frac{1}{\gamma}-\frac{1}{2}\ln \left(1+\frac{1}{\gamma^2}\right) \right]
        +2\re\left\{\left(1+i\gamma\right)\dilog\left(\frac{1}{1+i\gamma}\right)\right\}\\
        &
        -2\ln(1+r)
        \bigg[\,\frac{1+r}{r}-\gamma\atan\frac{1+r}{r\gamma}
          -\frac{1}{2}\ln\bigg(1+\left(\frac{1+r}{r\gamma}\right)^2\bigg)\bigg]\\
        &+2\re\left\{(1+i\gamma)\dilog\left(\frac{1}{1+r(1+i\gamma)}\right)\right\}
        -2\re\left\{(1+i\gamma)\dilog\left(\frac{1+r}{1+r(1+i\gamma)}\right)\right\}\;,
    \end{aligned}
\end{equation}
where $r=2p_iq_i/q_i^2$. The numerical evaluation of this expression is slow,
due to the various transcendental functions involved. In practice, we therefore perform the
one-dimensional integral in Eq.~\eqref{eq:finite_remainder_unintegrated} numerically.

\section{Validation and phenomenology}
\label{sec:pheno}
This section contains validation results as well as first phenomenological results based on the
parton shower algorithm described in Secs.~\ref{sec:resonance-aware} and~\ref{sec:width-aware},
the fixed-order subtraction described in Sec.~\ref{sec:nlo}, and a matching of the two using
the S-MC@NLO technique~\cite{Frixione:2002ik,Hoeche:2011fd,Hoeche:2012fm}.

\begin{table}[tp]
\begin{tabular}{p{5cm}|>{\raggedleft}p{15mm}|>{\raggedleft}p{15mm}|>{\raggedleft}p{15mm}|
>{\raggedleft}p{15mm}|>{\raggedleft}p{15mm}|>{\raggedleft}p{15mm}|>{\raggedleft\arraybackslash}p{15mm}}
  $e^+e^-\to W^+W^-b\bar{b}$,  &
  I [pb] & $\Delta$I [pb] & RS [pb] & $\Delta$RS [pb] & IRS [pb] & $\Delta$IRS [pb] & $\Delta$IRS/NLO \\
  $\sqrt{s}=365$ GeV, $\mu_R=m_t$ & & & & & & & \\\hline
  CS & 0.08453 & 0.00004 & -0.02185 & 0.00037 & 0.06268 & 0.00038 & 0.5\permil\\
  \Alaric default & 0.07293 & 0.00002 & -0.00997 & 0.00029 & 0.06296 & 0.00029 & 0.4\permil\\
  \Alaric res. aware & 0.13134 & 0.00003 & -0.06908 & 0.00048 & 0.06226 & 0.00048 & 0.6\permil\\
  \Alaric res. + width aware & 0.13472 & 0.00005 & -0.07248 & 0.00037 & 0.06224 & 0.00037 & 0.5\permil\\\hline
  CS-\Alaric default & \multicolumn{4}{c|}{} & \multicolumn{3}{c}{-0.00029 $\pm$ 0.00048}\\
  CS-\Alaric res. aware & \multicolumn{4}{c|}{} & \multicolumn{3}{c}{\phantom{-}0.00042 $\pm$ 0.00061}\\
  CS-\Alaric res. + width aware & \multicolumn{4}{c|}{} & \multicolumn{3}{c}{\phantom{-}0.00044 $\pm$ 0.00053}
\end{tabular}\vskip 5mm
\caption{Subtraction-scheme dependent cross-section contributions for $e^+e^-\to W^+W^-b\bar{b}$
  at $\sqrt{s}=365~\mathrm{GeV}$ with renormalization scale $\mu_R = m_t = 172.5~{\rm GeV}$.
  For numerical stability, we require a Durham jet distance of $y_{\rm 12}\, Q^2>1~{\rm GeV}^2$.\label{tab:xsecs}}
\end{table}
We first validate the correctness of the subtraction scheme by comparing inclusive cross-sections
for $e^+e^-\to W^+W^-b\bar{b}$ at $\sqrt{s}=365~{\rm GeV}$. We only present numbers
relevant for an informed comparison, i.e., the integrated counter-terms and the real-subtracted
cross-section for each subtraction scheme. The symbols in Tab.~\ref{tab:xsecs} are defined as follows:
$I$ indicates the contribution to the total NLO cross section from the integrated infrared subtraction
terms. $RS$ indicates the contribution to the total NLO cross section from the infrared subtracted
real-emission corrections. $IRS$ indicates the sum of integrated and real subtracted contributions,
i.e.\ $IRS=I+RS$. The results obtained with the default \Alaric kinematics and evolution kernel
without any modification, for the resonance aware kinematics with CS kernel as described in
Sec.~\ref{sec:resonance-aware} and, finally, with the width aware kernel from Eq.~\eqref{eq:soft-kernel}.
As an overall reference, we present numbers from the standard Catani-Seymour subtraction scheme
using the public \Sherpa version 3.0.3. We evaluate all contributions to a statistical precision
required to reach sub-permille level accuracy for NLO cross sections, and observe excellent agreement
between all subtraction schemes.

\begin{figure}
    \centering
    \includegraphics[width=0.49\linewidth]{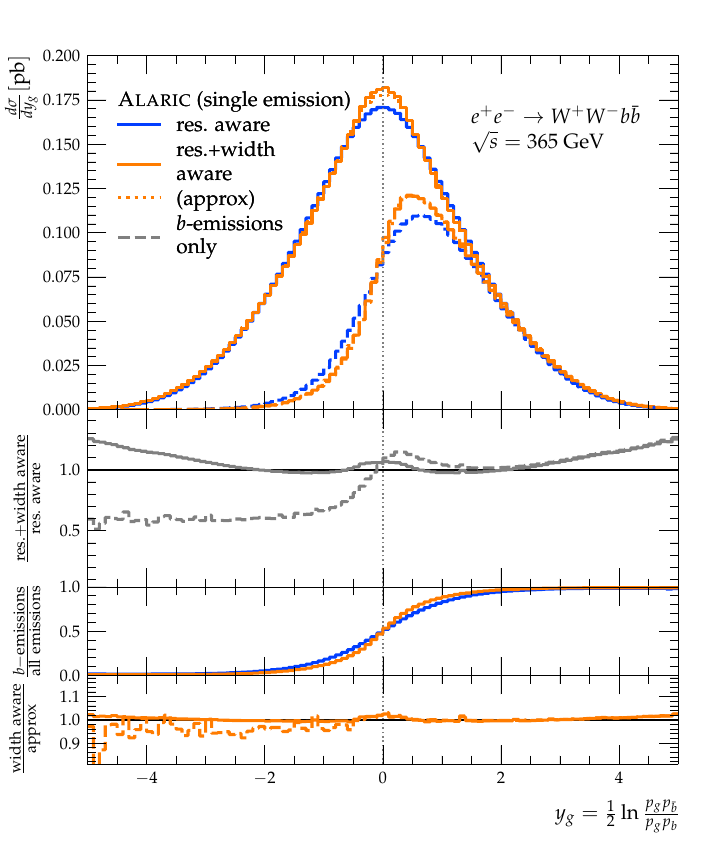}
    \includegraphics[width=0.49\linewidth]{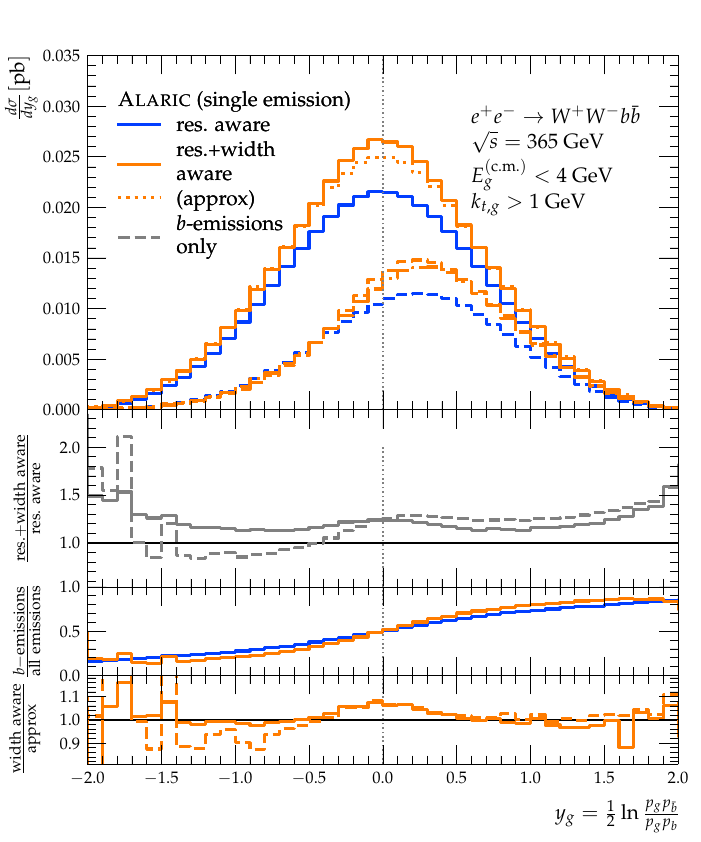}
    \caption{Comparison of Lund-plane rapidity of the first parton-shower emission in a standard
      evolution algorithm, and the resonance and width-aware approaches. Left: Full result.
      Right: Low-energy region, most affected by the width effects discussed in Sec.~\ref{sec:width-aware}.}
    \label{fig:1em-plots}
\end{figure}
As a first test of the QCD radiation pattern in the width-aware parton shower, we study the rapidity
of the first emitted particle in the Lund plane defined by the $b$ and the $\bar{b}$ quark
\begin{equation}
    y_g = \frac{1}{2}\ln \frac{p_gp_{\bar{b}}}{p_gp_b}\;.
\end{equation}
Emissions collinear to the $b$ quark will have large positive values of $y_g$, while emissions collinear
to the $\bar{b}$ quark will have large negative values. The left panel of Fig.~\ref{fig:1em-plots}
shows the distribution integrated over all gluon energies, while the right panel focuses on the
range $E_g^{\rm (c.m.)} < 4~{\rm GeV}$, where $\chi \lesssim 0.9$, using the approximation that
the top quarks are nearly at rest in the center of mass frame, cf.\ Sec.~\ref{sec:width-aware}.
For the right panel, we impose an additional cut on the gluon transverse momentum relative to
the $b\bar{b}$ dipole, $k_{t,g}>1~{\rm GeV}$. This eliminates regions of phase space, where
the interplay of the energy requirement and the shower cutoff conspire to allow emissions
collinear to the quark / antiquark only from soft wide angle radiation off the antiquark / quark.
Note that, with usual shower cutoffs $\sqrt{t_0}\approx 1~{\rm GeV}$ and the physical top width
$\Gamma_t \approx 1.3~{\rm GeV}$, there is no relevant region where $\chi \ll 1$ would hold,
and hence the radiation pattern of a standard $b\bar{b}$ dipole is never fully recovered.

The dashed histograms in Fig.~\ref{fig:1em-plots} display results obtained by selecting only gluons
radiated from the $b$ quark (selecting emissions from the $\bar{b}$ quark leads to identical plots
mirrored at $y_g=0$). The dotted histograms show results obtained with the parton-shower kernel
using the approximate expression $\chi(z)$ on the right hand side of Eq.~\eqref{eq:chi} instead
of the full expression on the left hand side in terms of the gluon energy in the frame of the top quark.

The full distribution, integrated over all energies, on the left of Fig.~\ref{fig:1em-plots},
shows an enhancement of the width aware evolution kernel in the central region, i.e. for small $|y_g|$.
However, emissions form the $b$-quark into the $\bar{b}$ hemisphere ($y_g<0$), are suppressed
in the width-aware evolution.
The approximation made in the implementation of $\chi$ has negligible effects, deviating from
the more accurate expression by at most $1\%$ and only at large rapidities. Accuracy in these
phase-space regions is less critical, because emissions in the collinear regime are identified
as part of the $b$-jet upon application of a jet algorithm. Considering only emissions from the
$b$ quark, the approximation holds only to a few per-cent in the opposite hemisphere, for $y_g<-1$,
leading to a somewhat stronger suppression in that region within the approximate treatment of $\chi$.

In the right panel of Fig.~\ref{fig:1em-plots}, which focuses on gluon energies around the top width,
we observe a similar pattern. Emissions from the $b$ quark into the opposite hemisphere are significantly
suppressed, but emissions in the $b$ hemisphere are enhanced also in the central region.
In the combination of emissions from both quarks, the emission probability is enhanced over the full
rapidity range, and relatively flat in $y_g$. The approximate treatment of $\chi$ leads to somewhat
larger effects here, as could have been anticipated since in this energy range $\chi$ interpolates
between the a parton shower with $bt$ and $\bar{b}\bar{t}$ dipoles, and one with a $b\bar{b}$ dipole.
However, the effects remain below $10\%$. For the overall distributions, this transition region
only plays a minor role, accounting for about $7\%$ of the cross section and contributing
soft gluons that will have a small effect on kinematical distributions.

\begin{figure}
    \centering
    \includegraphics[width=0.49\linewidth]{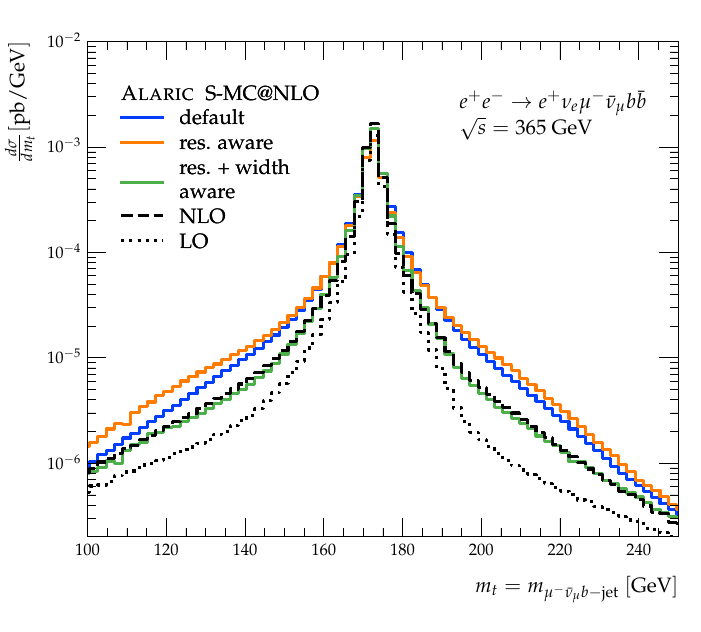}
    \includegraphics[width=0.49\linewidth]{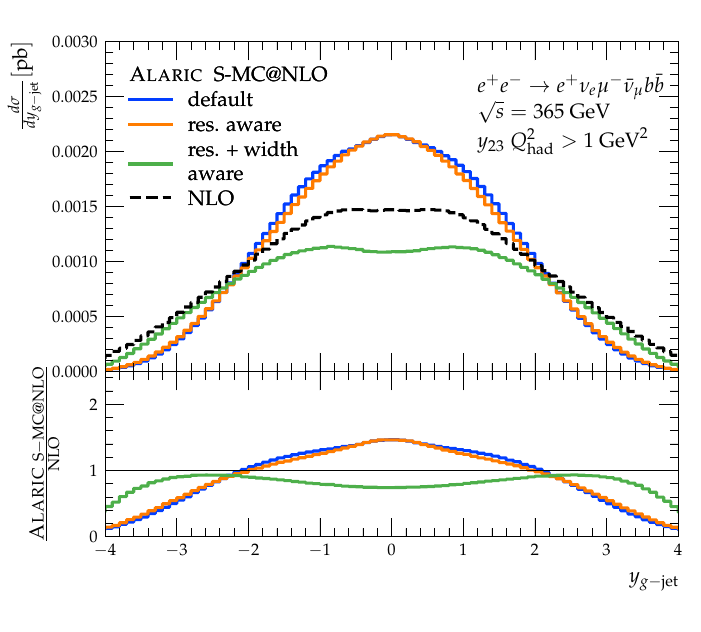}
    \caption{Comparison of NLO-matched predictions in the standard S-MC@NLO method and the
    resonance- and width-aware techniques described in Secs.~\ref{sec:resonance-aware}
    and~\ref{sec:width-aware}. $W$ bosons are reconstructed at truth level, including the neutrino momenta.
    Dotted and dashed lines show the fixed-order, leading order and next-to-leading order
    predictions, respectively. Left: Reconstructed top-quark mass. Right: Gluon-jet rapidity.}
    \label{fig:mcnlo-plots}
\end{figure}
We continue our analysis with NLO matched distributions at parton level. For an easy identification
of the final decay products of the top quarks, we focus on di-leptonic decays into different leptons. 
Specifically, we let the the $W^-$ decay to $\mu^-\bar{\nu}_\mu$ and the $W^+$ to $e^+\nu_e$.
The hadronic final-state particles are clustered into jets using the Durham algorithm \cite{Catani:1991hj}. 
We cluster all events into exactly 2 jets, and separately events containing at least 3 partons into exactly 3 jets.
In both cases, we discard events where the correspondence of the jets to 
decayed top quarks is ambiguous, i.e. we require that both (or exactly 2 of the 3) jets contain
exactly one $b$ quark, and that those $b$ quarks are of opposite charge. This means that both
$b$-jets can be combined with the corresponding $W$ boson to form a reconstructed top/anti-top quark,
while the third jet in the 3-jet case acts as a proxy for the emitted gluon. 
The third jet is only considered if the relevant clustering distance was $y_{23}\, Q_{\rm had}^2 > 1~{\rm GeV}^2$. 
We first analyze events with truth level reconstructed $W$ bosons, i.e. including the neutrino momenta.
In Fig.~\ref{fig:mcnlo-plots}, we show the top-quark mass reconstructed from the 2-jet clustering,
and the rapidity distribution of the gluon-proxy jet. In the latter case we measure the rapidity
relative to the directions of the $b$-jets. Both the default and the width aware matching scheme
predict a significant enhancement of the tails of the top-quark mass distribution. 
Since the resonance aware shower does by construction not change the total mass of the top decay products
including radiation off of them, this change is due to clustering mismatches where radiation
from the $b$-quark is assigned to the jet containing the $\bar{b}$ quark and vice-versa.
The gluon-jet rapidity distribution confirms that indeed the first two matching schemes
predict significantly more radiation in the central region, while the fixed order result
shows a marked depletion, consistent with the expectation that, in most of the relevant phase space,
both $b$-quarks radiate gluons independently, rather than as a coherent dipole. 
Only the width aware parton shower preserves this behavior after the matching.
\begin{figure}
    \centering
    \includegraphics[width=0.49\linewidth]{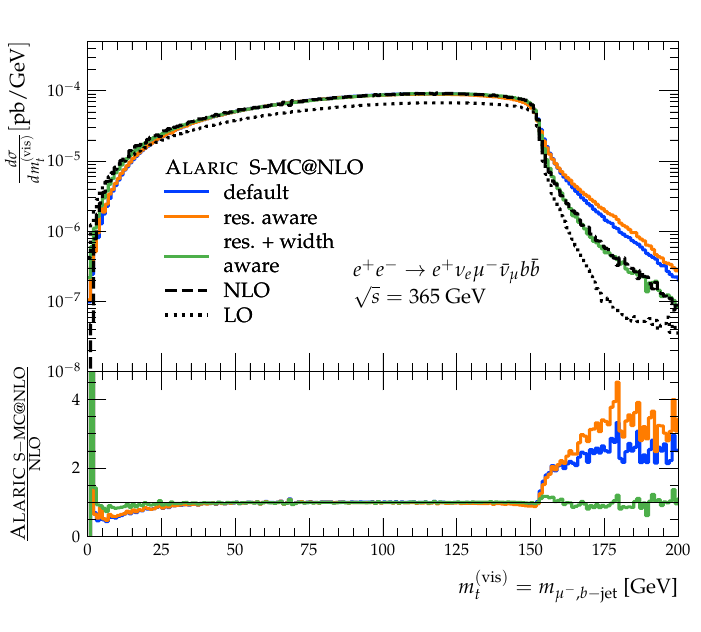}
    \includegraphics[width=0.49\linewidth]{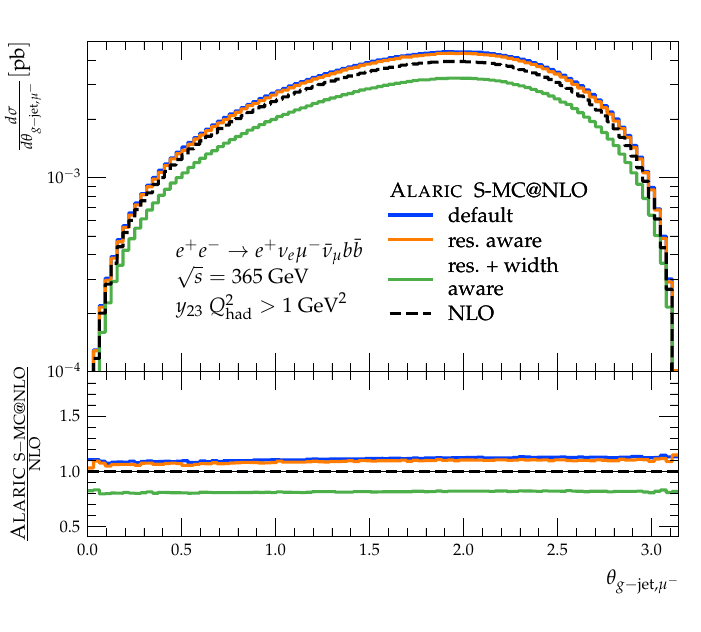}
    \caption{Comparison of NLO-matched predictions in the standard S-MC@NLO method and the
    resonance- and width-aware techniques described in Secs.~\ref{sec:resonance-aware}
    and~\ref{sec:width-aware}. $W$ bosons decay leptonically.
    Dotted and dashed lines show the fixed-order, leading order and next-to-leading order predictions, respectively.
    Left: Visible anti-top quark mass. Right: Angular separation between gluon jet and lepton.}
    \label{fig:mcnlo-plots-2}
\end{figure}

We finally discuss a setup where the leptonic decays of the $W$ bosons are taken into account. 
The left panel of Fig.~\ref{fig:mcnlo-plots-2} shows the visible mass reconstructed
from $\mu^-$ and the corresponding $b$-jet. The right panel shows the angular
separation between the gluon jet and the $\mu^-$. We observe that the width aware
splitting kernels produce fewer additional jets than both the default and the
resonance aware matched showers, and also slightly fewer than predicted at fixed order.
The angular separation is otherwise unaffected by the matching,
and the shape matches the fixed order result in all three cases.
The visible top mass shown in the left panel displays the peak structure typical of
such an observable. The large matching effects for the default and resonance aware
matching schemes already observed in Fig.~\ref{fig:mcnlo-plots} are concentrated in the
high-mass tail of Fig.~\ref{fig:mcnlo-plots-2} left, which is known to be dominated
by the shape of the Breit-Wigner distribution of the resonance virtuality.
Even though the neutrino is consistently treated as missing energy, there still
remains a factor $> 2$ difference between resonance aware and width-aware matching
as a consequence of the effects discussed in Sec.~\ref{sec:width-aware}.
Only the parton shower matched in the width aware scheme closely follows
the fixed-order prediction in that region. Given the expected statistical
accuracy and systematic precision of a potential FCC-ee, the differences between
resonance- and width-aware evolution would impact potential template fits performed
on this observable.

\section{Outlook}
\label{sec:outlook}
Standard parton shower simulations do not account for potential resonances appearing
in the hard process when simulating the radiation of soft gluons from external legs. 
This simplification is insufficient for several reasons. Firstly, additional radiation
can significantly alter the propagator structure of the hard process and shift narrow
resonances off their mass shell, which poses problems for infrared subtraction and the
associated parton shower matching. Secondly, the radiation from color charged particles
connected by resonances is often significantly altered by nontrivial width effects.
They impact only energy scales above the width, and will therefore not be described
correctly by a formalism that determines the radiation pattern purely based on the
infrared limit of the scattering amplitudes. However, due to its higher energy, 
the radiation at scales above the width of resonances can also be far more important
in practice  than the radiation produced at lower scales.

In this manuscript, we have presented an approach that addresses the challenges associated
with internal resonances, using a modern parton shower with NLL-preserving kinematics mapping.
In the context of $t\bar{t}$ production at the FCC-ee, we introduced a fully resonance aware
evolution and subtraction scheme based on the \Alaric method.

Our method to achieve resonance awareness in the parton shower and matching is general,
and can be applied to processes like top quark production (and subsequent decays) at the LHC.
Refinements in the strategy for resonance identification will, however, be necessary for
processes with competing resonances. Regarding the physics modeling of color singlet
$t\bar{t}$ production, the main restriction currently appears to be the requirement to be
close to threshold, such that terms of $\mathcal{O}(v_t)$ can be neglected, but not close
enough for Coulomb effects of $\mathcal{O}(\alpha_s/v_t)$ to play a role. 
We performed our analysis in the phenomenologically most relevant region for FCC-ee,
at $\sqrt{s}=365~{\rm GeV}$, where both these effects are suppressed. An improved calculation
should include resummation of Coulomb effects (similar to Ref. \cite{Bach:2017ggt}),
a full NLO fixed order treatment of the top decays, and effects of QED radiation.

\section*{Acknowledgments}
\noindent
This manuscript has been authored by Fermi Forward Discovery Group, LLC
under Contract No. 89243024CSC000002 with the U.S.\ Department of Energy,
Office of Science, Office of High Energy Physics.
This research used resources of the National Energy Research Scientific Computing Center (NERSC), 
a Department of Energy Office of Science User Facility using NERSC award ERCAP0028985.
The work of S.H. was supported by the U.S. Department of Energy,
Office of Science, Office of Advanced Scientific Computing Research,
Scientific Discovery through Advanced Computing (SciDAC-5) program,
grant “NeuCol”. D.R.\ is supported by the European Union under the HORIZON program in
Marie Sk{\l}odowska-Curie project No. 101153541. During parts of this work, D.R. 
was further supported by STFC under grant agreement ST/P006744/1 as well as 
by the Durham University Physics Department Developing Talents Award and by 
the German Academic Exchange Service (DAAD).

\appendix

\section{Alaric Kinematics}
\label{app:kinematics}
\begin{figure}[ht]
  \centerline{\includegraphics[width=0.9\textwidth]{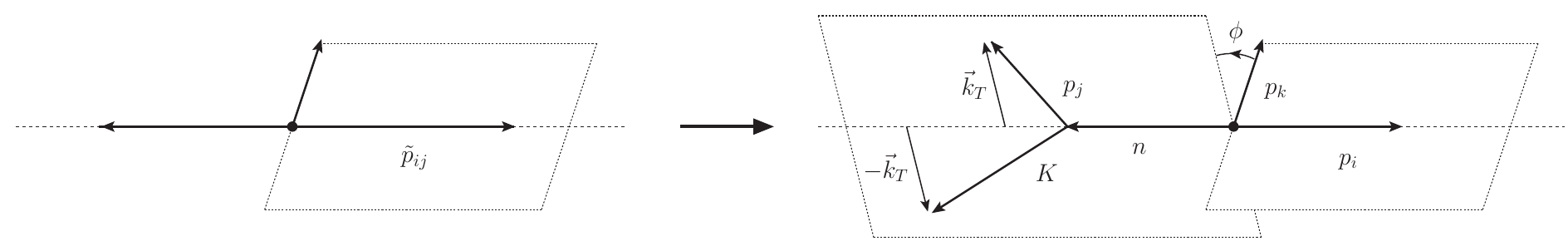}}
  \centerline{\includegraphics[width=0.9\textwidth]{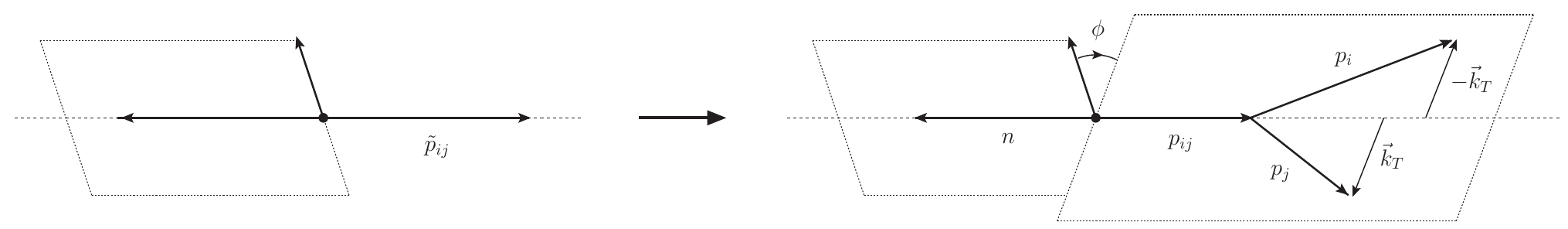}}
  \caption{Top: The radiation kinematics, described by Eqs.~\eqref{eq:fi_emit_spec}.
    The momentum $K$ takes the transverse recoil in the splitting, and the momentum
    $p_k$ defines a reference direction for the azimuthal angle.
    Bottom: The splitting kinematics, described by Eqs.~\eqref{eq:fi_emit_emiss}.
    The momentum $p_i$ takes the transverse recoil in the splitting, and the momentum
    $p_k$ defines a reference direction for the azimuthal angle.
    \label{fig:kinematics}}
\end{figure}
Following Refs.~\cite{Assi:2023rbu,Hoche:2025anb}, we define two different types
of momentum mapping. These parametrizations are designed to reflect the dynamics
of the radiation and decay components of the splitting functions.
The radiation kinematics, used in conjunction with scalar (soft enhanced)
components of the splitting functions, are sketched in Fig.~\ref{fig:kinematics}~(top).
The phase space is parameterized in terms of the variables~\cite{Catani:1996vz,Herren:2022jej}
\begin{equation}\label{eq:def_v_z_cs}
  v=\frac{p_ip_j}{p_i\tilde{K}}
  \qquad\mathrm{and}\qquad
  z=\frac{p_i\tilde{K}}{\tilde{p}_i\tilde{K}}\;.
\end{equation}
For massless partons, the final-state momentum of the emitter, $\tilde{p}_i$,
and the recoil momentum, $\tilde{K}$, are given by
\begin{equation}\label{eq:fi_emit_spec}
  \begin{split}
    p_i=&\;\,z\,\tilde{p}_i\;,\\
    p_j=&\;\,(1-z)\,\tilde{p}_i+v\big(\tilde{K}-(1-z+2\kappa)\,\tilde{p}_i\big)-k_\perp\;,\\
    K=&\;\tilde{K}-v\big(\tilde{K}-(1-z+2\kappa)\,\tilde{p}_i\big)+k_\perp\;,
  \end{split}
\end{equation}
with the magnitude squared of the transverse momentum defined as
\begin{equation}
  {\rm k}_\perp^2=v(1-v)(1-z)\,2\tilde{p}_i\tilde{K}-v^2\tilde{K}^2\;.
\end{equation}
The vector $n$ in Fig.~\ref{fig:kinematics} is defined as $n=K+p_{j}$.
If the recoil momentum $\tilde{K}$ is composed of multiple momenta (e.g.\ because
the recoil is distributed onto a system of multiple particles), the individual momenta
are subjected to a Lorentz transformation
\begin{equation}\label{eq:lorentz_trafo_fs}
  p_l^\mu\to\Lambda^\mu_{\;\nu}(K,\tilde{K}) \,p_l^\nu\;,
  \qquad\text{where}\qquad
  \Lambda^\mu_{\;\nu}(\tilde{K},K)=g^\mu_{\;\nu}
  -\frac{2(K+\tilde{K})^\mu(K+\tilde{K})_\nu}{(K+\tilde{K})^2}
  +\frac{2K^\mu \tilde{K}_\nu}{\tilde{K}^2}\;.
\end{equation}
The splitting kinematics, used in conjunction with non-scalar (non-soft enhanced)
components of the splitting functions, is sketched in Fig.~\ref{app:kinematics}~(bottom).
We make use of some of the notation in~\cite{Catani:2002hc}, in particular
\begin{equation}\label{eq:def_y_z_cdst}
    y=\frac{p_ip_j}{p_ip_j+p_iK+p_jK}\;,
    \qquad
    z=\frac{p_iK}{p_iK+p_jK}\;,
\end{equation}
and we define the scaled recoil mass $\kappa=\tilde{K}^2/(2\tilde{p}_{ij}\tilde{K})$.
the momenta after the splitting are given by
\begin{equation}\label{eq:fi_emit_emiss}
  \begin{split}
    p_i^\mu=&\;\bar{z}\;
    \frac{\tilde{p}_{ij}^\mu}{z_{ij}}+y\,(1-\bar{z})\,
    z_{ij}\big(\tilde{K}^\mu-\kappa\,\tilde{p}_{ij}^\mu\big)
    +k_\perp^\mu\;,\\
    p_j^\mu=&\;(1-\bar{z})\;
    \frac{\tilde{p}_{ij}^\mu}{z_{ij}}+y\,\bar{z}\,
    z_{ij}\big(\tilde{K}^\mu-\bar{\kappa}\,\tilde{p}_{ij}^\mu\big)
    -k_\perp^\mu\;,
  \end{split}
\end{equation}
where
\begin{equation}
  z_{ij}=\frac{1+y-\sqrt{(1-y)^2-4y \kappa}}{2y(1+\kappa)}\;,
  \qquad
  \bar{z}=\frac{\displaystyle zz_{ij}(1-y)(1-z_{ij}y)
  -z_{ij}^2y\kappa}{(1-z_{ij}y)^2-z_{ij}^2y\kappa}\;,
  \qquad
  {\rm k}_\perp^2=2\tilde{p}_{ij}\tilde{K}\,\bar{z}(1-\bar{z})\alpha_{ij}\;.
\end{equation}
The construction of the transverse momentum vector, $k_\perp^\mu$, and the
phase-space factorization are discussed in Ref.~\cite{Assi:2023rbu}.

\section{Soft integrals}
\label{app:soft-integrals}
In this appendix we re-derive the integral of the scalar component of the splitting
functions, $V^{(s,CS)}_{ig,k}$, in Eq.~\eqref{eq:eik_partfrac_cs}, using the notation
of Ref.~\cite{Assi:2023rbu}. In radiation kinematics, with $\tilde{K}=p_W$, we find
the following expression for the one-emission phase space:
\begin{equation}\label{eq:ps_one_emission}
    \begin{aligned}
    d\Phi_{+1} &= \left(\frac{2\tilde{p}_i\tilde{K}}{16\pi^2}\right)^{1-\eps} \frac{(z(1-z))^{1-2\eps}}{(1-z+\kappa)^{1-\eps}} {\rm d}z \frac{{\rm d}\Omega^{2-2\eps}}{4\pi} \\
    &= (2p_ip_k)^{-\eps} \frac{1}{(16\pi^2)^{1-\eps}} \left(\frac{np_k}{np_i}\right)^\eps\left(\frac{l_{ik}^2}{4}\right)^\eps z^{1-\eps} \frac{(2\tilde{p}_i\tilde{K})^2}{n^2}\frac{{\rm d}z}{(1-z)^{-1+2\eps}}\frac{{\rm d}\Omega^{2-2\eps}}{4\pi}\;.
    \end{aligned}
\end{equation}
The generic form of the partial fractioned eikonal is taken from Ref.~\cite{Catani:1996vz}.
It reads
\begin{equation}
    \langle V^{(s,CS)}_{ij,A} \rangle = 8\pi\mu^{2\eps}\alpha_s \frac{2\,p_ip_A}{p_ip_j+p_Ap_j}\;.
\end{equation}
Combining this with the differential phase-space element in Eq.~\eqref{eq:ps_one_emission},
we obtain
\begin{equation}
    \begin{aligned} 
        d\Phi_{+1} \frac{1}{2p_ip_j}\langle V^{(s,{\rm A})}\rangle &= \frac{\alpha_s}{2\pi} \left(\frac{4\pi\mu^2}{2p_ip_k}\right)^{\eps} (4\pi)^{\eps} \left(\frac{np_k}{np_i}\right)^\eps\left(\frac{l_{ik}^2}{4}\right)^\eps  \frac{z^{1-\eps}\,{\rm d}z}{(1-z)^{-1+2\eps}}\frac{{\rm d}\Omega^{2-2\eps}}{4\pi}   \frac{(2\tilde{p}_i\tilde{K})^2}{2p_ip_j\,n^2}\frac{2\,p_ip_A}{p_ip_j+p_Ap_j}\\
        &= S_\eps \left(\frac{np_k}{np_i}\right)^\eps\left(\frac{l_{ik}^2}{4}\right)^\eps \frac{1}{2\pi}\frac{\Gamma^2(1-\eps)}{\Gamma(1-2\eps)} \frac{z^{1-\eps}\,{\rm d}z}{(1-z)^{-1+2\eps}}\frac{{\rm d}\Omega^{2-2\eps}}{\Omega(1-2\eps)}   \frac{(2\tilde{p}_i\tilde{K})^2}{2p_ip_j\,n^2}\frac{2\,p_ip_A}{p_ip_j+p_Ap_j}\;,
    \end{aligned}
\end{equation}
where the following quantities have been defined:
\begin{equation}
  S_\eps = \frac{\alpha_s}{2\pi} \left(\frac{4\pi\mu^2}{2p_ip_k}\right)^{\eps} \frac{1}{\Gamma(1-\eps)}\;,
  \qquad\text{and}\qquad
  \Omega(1-2\eps) = \frac{\Gamma(1-\eps)}{\Gamma(1-2\eps)} \frac{2}{(4\pi)^\eps}\;.
\end{equation}
Note that the apparent dependence of the result on $k$ has been introduced
in the prefactors only by factoring out the conventional $(2p_ip_k)^{-\eps}$.
We can rewrite the remaining integral as follows: 
\begin{equation}
  \int \frac{{\rm d}\Omega^{2-2\eps}}{\Omega(1-2\eps)}
  \frac{(2\tilde{p}_i\tilde{K})^2}{2p_ip_j\,n^2}\frac{2\,p_ip_A}{p_ip_j+p_Ap_j}
  = \frac{4}{(1-z)^2} l_il_{i+A}I^{(1)}_{1,1}(l_il_{i+A},l_{i+A}^2)\;.
\end{equation}
The scalar two-mass angular integral on the right-hand side is given by~\cite{
  vanNeerven:1985xr,Beenakker:1988bq,Somogyi:2011ir,Isidori:2020acz,Lyubovitskij:2021ges} 
\begin{equation}
  I_{1,1}^{(1)}(v_{12},v_{11}) =
  -\frac{\pi}{v_{12}}\left(\frac{v_{11}}{v_{12}^2}\right)^\eps
  \left[\,\frac{1}{\eps} + 2 \eps
  \left(\dilog(1-\frac{v_{12}}{1-\sqrt{1-v_{11}}}) + \dilog(1+\frac{v_{12}}{1-\sqrt{1-v_{11}}})\right)
  \right]\;,
\end{equation}
where
\begin{equation}
        l_il_{i+A} = \frac{(p_ip_A) n^2}{np_i (np_i+np_A)}\;,
        \qquad
        l_{i+A}^2  = \frac{(2p_ip_A+p_A^2) n^2}{(np_i+np_A)^2}\;,
        \qquad\text{and}\qquad
        l_{iA}^2   = \frac{(2p_ip_A+p_A^2)n^2}{(np_i)(np_A)}\;.
\end{equation}
The final expression for the integrated infrared subtraction term is
\begin{equation}
    \int d\Phi_{+1} \frac{1}{2p_ip_j}\langle V^{(s,{\rm A})}\rangle = S_\eps \int \left(\frac{np_k}{np_i}\right)^\eps\left(\frac{l_{ik}^2}{4}\right)^\eps \frac{2}{\pi}\frac{\Gamma^2(1-\eps)}{\Gamma(1-2\eps)} \frac{z^{1-\eps}\,{\rm d}z}{(1-z)^{1+2\eps}}\, l_il_{i+A}I^{(1)}_{1,1}(l_il_{i+A},l_{i+A}^2)\;,
\end{equation}
which is typically evaluated as a Laurent series in $\eps$ around the soft pole:
\begin{equation}
  \begin{aligned}
  I^{(s,\text{CS})}_{ik} &= \int_0^1 \mathop{\mathrm{d}z}
    \left(-\frac{\delta(1-z)}{2\eps}+\frac{z}{\left[1-z\right]_{+}}
    - 2\eps z \left[\frac{\log (1-z)}{1-z}\right]_{+} \right)\\
    &\phantom{= \int_0^1 \mathop{\mathrm{d}z} } \times \frac{2}{\pi} z^{-\eps}
    \left(\frac{np_k}{np_i} \frac{l_{ik}^2}{4}\right)^{\eps}\frac{\Gamma^2(1-\eps)}{\Gamma(1-2\eps)}\,
     l_il_{i+A}I^{(1)}_{1,1}(l_il_{i+A},l_{i+A}^2)~.
  \end{aligned}
\end{equation}

\FloatBarrier

\bibliography{main}
\end{document}